\documentclass[runningheads]{llncs}
\usepackage{graphicx}

\usepackage{bm}
\usepackage{booktabs}
\usepackage{subcaption}
\usepackage{verbatimbox}
\usepackage{amssymb}

\usepackage{custom}
\let\citet\cite

\newif\ifExt
\Exttrue

\begin{document}

\title{EqFix: Fixing \LaTeX~Equation Errors by Examples\thanks{
  This work was supported in part by the National Natural Science Foundation of China (No. 62072267 and No. 62021002) and the National Key Research and Development Program of China (No. 2018YFB1308601).
}}

\author{
  Fengmin Zhu\inst{1,2}\thanks{Early revisions of this work were done when this author was in Tsinghua University.} \and
  Fei He$^{\textrm{(\,\Letter\,)}}$\inst{1}
}
\authorrunning{F. Zhu et al.}

\institute{Tsinghua University \\ \email{hefei@tsinghua.edu.cn} \and
Max Planck Institute for Software Systems}

\maketitle

\begin{abstract}
  \LaTeX{} is a widely-used document preparation system.
Its powerful ability in mathematical equation editing is perhaps the main reason for its popularity in academia.
Sometimes, however, even an expert user may spend much time fixing an erroneous equation.
In this paper, we present \EqFix, a synthesis-based repairing system for \LaTeX{} equations.
It employs a set of fixing rules and can suggest possible repairs for common errors in \LaTeX{} equations.
A domain-specific language is proposed for formally expressing the fixing rules.
The fixing rules can be automatically synthesized from a set of input-output examples.
An extension of relaxers is also introduced to enhance the practicality of \EqFix.
We evaluate \EqFix on real-world examples and find that it can synthesize rules with high generalization ability.
Compared with a state-of-the-art string transformation synthesizer,
\EqFix solved 37\% more cases and spent less than half of their synthesis time.

\end{abstract}

\section{Introduction}

\LaTeX{} is a text-based document preparation system widely used in academia to publish and communicate scientific documents.
The powerful typesetting of mathematical equations makes it a universal syntax for expressing mathematical equations.
This syntax has been integrated into text-based markup languages like
Markdown\footnote{\url{https://daringfireball.net/projects/markdown/}},
and WYSIWYG (\ie ``What you see is what you get'') document processors like
MS Word\footnote{\url{https://products.office.com/en-us/word}}.
Even on the web, one can display \LaTeX{} equations beautifully by MathJax\footnote{\url{https://www.mathjax.org}}.

Since the syntax of \LaTeX{} equations is quite \emph{complex}, \emph{non-expert} users may find it challenging to use.
For example, one may expect ``$x^{10}$'' by typing ``\verb|$x^10$|''.
This equation indeed compiles; however, its actual output is ``$x^10$'', which goes against the user's expectation.
When such an error occurs, one may resort to online help forums.
However, this process is \emph{never trivial}.
First, the user has to provide several keywords (for searching) that well describe the error.
Second, even if the user is fortunate to obtain some solutions, they may not necessarily work for the user's problem---the user still has to adapt the answers to that specific problem.
The whole process---especially the adaption of the solutions to one's own problem---requires not only intelligence but also patience.

\emph{Programming by examples} (PBE) is believed to be a revolutionary technique for end-user programming \cite{pbe,flash-fill,flash-normalize,flash-extract,ringer}.
In recent decades, PBE has been adopted in the area of \emph{program repair} \cite{PatchGen,S3}.
\NoFAQ \cite{NoFAQ} is a tool that employs \emph{error messages} to assist the PBE-based repairing.
This tool aims to fix common errors in Unix commands, from an input-output example that consists of an erroneous Unix command, an error message prompted by Shell, and a rectified command specified by experts.
Note that the error messages prompted by Unix Shell are usually instantiated from a set of predefined templates;
the stored information in the messages can thus be easily extracted by patterns.

Inspired by \NoFAQ and other PBE techniques, we present \EqFix, a system for automatically fixing erroneous \LaTeX{} equations by examples.
Note that \NoFAQ cannot be applied to our problem setting for two reasons:
First, an equation error is not necessarily a compilation error, such as ``\verb|$x^10$|'' indeed compiles but produces an unexpected result.
For such errors, one needs to specify the error message on their own.
Thus, we cannot assume this message has a fixed structure (\ie is instantiated from a template) as in \NoFAQ.
Second, the Unix command can be directly \emph{tokenized} into a sequence of strings (using whitespaces as the delimiters), which makes it straightforward (by comparing the tokenized strings in turn) to locate the error position in the text of the Unix command.
However, it is not the same case for equation text.
Instead, we have to collaborate the corresponding \emph{error message} to tokenize an equation.

To the best of our knowledge, \EqFix is the first attempt at equation repair using PBE techniques.
Novice users can use it to automatically fix common errors in \LaTeX{} equations; expert users can contribute corrections for erroneous equations.
We design a \emph{domain-specific language} (DSL) for formally defining the fixing rules for erroneous \LaTeX{} equations 
(\cref{sec:lang}).
Intuitively, a fixing rule consists of 
an \emph{error pattern}, which specifies what error messages this rule is
applicable, and a \emph{transformer} which performs the actual fixing via string transformation. 
We propose an algorithm for synthesizing fixing rules expressed by our DSL
from input-output examples (\cref{sec:syn}).
We also introduce \emph{relaxers} to describe the \emph{generalization} of equation patterns, with which the search space relating to the faulty parts is expanded so that we can handle more repairing problems.

We evaluated \EqFix on a dataset containing 89 groups of real-world examples (\cref{sec:eval}).
Note that \NoFAQ is limited to repairing buggy Unix commands; we instead took the state-of-the-art PBE tool \FlashFill as our baseline.
We selected the longest example in each group as our test case.
We found that \EqFix solved 72 (80.9\%) of the test cases in less time, whereas \FlashFill solved only 39 (43.82\%).
Our prototype implementation and experiment artifact are publicly available: \url{https://github.com/thufv/EqFix}.

The main technical contributions of this paper are summarized as follows:
\begin{itemize}
    \item We present \EqFix, a PBE-based system for automatically fixing erroneous \LaTeX{} equations.
    \item We design a DSL for formally expressing fixing rules.
    We rely on equation patterns to extract and transfer relevant information between error messages
    and equations.
    The patterns can be generalized by relaxers when necessary.
    \item We conducted experiments on real-world examples.
    Results reveal the high effectiveness and applicability of our approach.
\end{itemize}

\section{EqFix by Examples}\label{sec:motivating}

We use several real-world examples (in \cref{tb:motivating}) to showcase how \EqFix repairs \LaTeX{} equations.
All of the examples were extracted from an online \LaTeX{} forum\footnote{\url{https://tex.stackexchange.com}}.
To be clear and short, we neglect the unchanged substrings of long equations.
Each example consists of three components -- the input equation $eq$, the error message $err$, and the fixed equation $fix$.
For convenience of reference, we number these examples from 1 to 8 and refer their components as $eq_i$, $err_i$, $fix_i$, for $1 \le i \le 8$, respectively.
The \LaTeX{} output on each equation (if it compiles) is displayed below the equation text.

\begin{table*}[t]
    \centering\footnotesize
    \caption{Motivating examples.}
    \label{tb:motivating}
    \begin{tabular}{llll}
        \toprule
        \# & $eq$ & $err$ & $fix$ \\ 
        \midrule
        1 & \verb|$x^10$| 
          & \verb|superscript 10| 
          & \verb|$x^{10}$| 
          \\
          & $x^10$ & & $x^{10}$ \\
        2 & \verb|$y^123+x$| 
          & \verb|superscript 123|
          & \verb|$y^{123}+x$| 
          \\
          & $y^123+x$ & & $y^{123}+x$ \\
        3 & \verb|$f^(k)$|
          & \verb|superscript (k)|
          & \verb|$f^{(k)}$| 
          \\
          & $f^(k)$ & & $f^{(k)}$ \\
        4 & \verb|$y=x+\ldots+x^10$|
          & \verb|superscript 10|
          & \verb|$y=x+\ldots+x^{10}$| 
          \\
          & $y=x+\ldots+x^10$ & & $y=x+\ldots+x^{10}$ \\
        \midrule
        5 & \verb|${1,2,3$| 
          & \verb|Missing } inserted|
          & \verb|${1,2,3}$| 
          \\
          & & & ${1,2,3}$ \\
        6 & \verb|$S={x_1,\ldots,x_n$| 
          & \verb|Missing } inserted|
          & \verb|$S={x_1,\ldots,x_n}$|
          \\
          & & & $S={x_1,\ldots,x_n}$ \\
        \midrule
        7 & \verb|$2\^x$|
          & \verb|Command \^ invalid|
          & \verb|$2^x$| 
          \\
          && \verb|in math mode| & $2^x$ \\
        8 & \verb|$\sum\limits_{i=1}\^N t_i$|
          & \verb|Command \^ invalid|
          & \verb|$\sum\limits_{i=1}^N t_i$|
          \\
          && \verb|in math mode| & $\sum\limits_{i=1}^N t_i$ \\
        \bottomrule
    \end{tabular}
\end{table*}

\begin{table}
    \centering\small
    \caption{Selected keywords supported by \EqFix.}
    \label{tb:keywords}
    \begin{tabular}{ll}
        \toprule
        Keywords  & Interpretations               \\ 
        \midrule
        superscript & expected as a superscript \\
        subscript   & expected as a subscript   \\
        set         & expected as a set \\
        function    & expected as a math function/operator \\
        greek letter    & expected as a greek letter \\
        fraction numerator   & expected as the numerator of a fraction \\
        fraction denominator & expected as the denominator of a fraction \\
        operator sum         & expected as a sum operator \\
        operator product     & expected as a product operator \\
        long arrow           & expected as a long arrow \\
        \bottomrule
    \end{tabular}
\end{table}

Examples \#1 -- \#4 present a scene where a user expects a superscript but forgets to parenthesize the superscript expression.
As shown in \#1, given the input equation ``\verb|$x^10$|'' ($eq_1$),
\LaTeX{} treats only ``\verb|1|'' but not the entire number ``\verb|10|'' as the superscript.
In this way, it outputs ``$x^10$'', which is against the user's intent.
Note that the input equation $eq_1$ itself is syntactically correct because the \LaTeX{} compiler did not report any error.
In this way, the user must specify an error message by hand to express their intent.
To express the error type conveniently, a set of predefined \emph{keywords} (\cref{tb:keywords} presents a selected subset) are provided for selection.
In a future direction, via natural language processing, we may simply accept a natural language sentence as the error message for even better practicality.
Then, the user needs to point out a substring of the erroneous input equation to show the error location.
The error message is a combination of the keywords and substring of the input equation.
Here, in $err_1$, ``\verb|superscript|'' is a predefined keyword indicating some substring of the input equation is expected to be the superscript, and ``\verb|10|'' gives the error location.
Then, an expert may fix\footnote{
    As another option, we may get the fix by online search.
} the input equation as ``\verb|$x^{10}$|'', \ie surrounding ``\verb|10|'' with a pair of curly brackets.
The three components, \ie $(eq_1, err_1, fix_1)$, compose an \emph{input-output example}, with which we can synthesize a \emph{(fixing) rule}.
Each rule consists of an \emph{error pattern} for matching the error message and a \emph{transformer} that will be applied to the input equation to produce a fix.
Intuitively, the underlying fixing strategy of this rule would be ``surrounding the superscript with a pair of curly brackets''.

\EqFix can switch between the \emph{training mode} for synthesizing rules and the \emph{applying mode} for repairing erroneous equations, based on a \emph{rule library} that saves all the learned rules so far.
In the training mode, \EqFix takes user-given examples (typically by expert users) as input.
It first searches in its rule library to obtain a rule that can be \emph{refined} to be \emph{consistent} with the new examples.
For instance, the example \#2 can be added by refining the fixing rule synthesized merely by \#1. 
If it is not the case (for instance, consider the examples \#5 to \#8), a new rule is synthesized and the rule library gets enlarged.

In the applying mode, \EqFix attempts to solve an \emph{equation repair problem}---an erroneous equation together with an error message---typically provided by an end user.
To do so, it searches in its rule library for all applicable rules, \ie those whose error patterns can match against the error message, and attempts to apply them (the transformer of the rule) to the input equation.
For instance, the rule synthesized from examples \#1 and \#2 is applicable to equation repair problems \#3 and \#4: applying this rule on $(eq_3, err_3)$ and $(eq_4, err_4)$ gives $fix_3$ and $fix_4$ respectively.
Since there can be more than one applicable rule, users are asked to review the suggested fixes and approve one that meets the intent.
If no rule is applicable, or all suggested fixes are rejected by the user, \EqFix fails on this equation repair problem.
In that situation, we expect an expert user to figure out a correction, which, in association with the erroneous equation and the error message, forms a new example for synthesis under the training mode.
The newly synthesized rule will be recorded in the rule library so that it can apply (under the applying mode) to future equation repair problems in this category.

Sometimes, an erroneous equation contains multiple errors.
One needs to interact with \EqFix in multiple rounds to fix them all.
The user feeds the erroneous equation together with one of the error messages in the initial round
and iteratively corrects the other errors using the fixed equation of the last round.

The rest examples in \cref{tb:motivating} showcase two \LaTeX compile errors: unmatched brackets (\#5 -- \#6) and invalid superscript operator (\#7 -- \#8).
The error messages prompted by the \LaTeX compiler are instantiated from some templates defined by \LaTeX.
Both compiler-prompted and keywords-based error messages are handled in a unified way (we will explain that in \cref{sec:error-pattern}).
Back to the examples, a possible correction for \#5 (\#6 is similar) suggested by an expert is ``\verb|${1,2,3}$|'', which inserts the missing right curly bracket (`\verb|}|') at the end of the equation.
A possible correction for \#7 (and \#8) is to use `\verb|^|'' in place of the erroneous ``\verb|\^|''.
Note that \#8 is more complicated than \#7, while it can be automatically fixed using the rule synthesized from \#7.

In summary, \EqFix facilitates an automated approach for fixing common errors in editing \LaTeX{} equations.
Our approach is rule-based (\cref{sec:lang}) and the synthesis by input-output examples (\cref{sec:syn}) is automated.
One benefit of our system is that we may collect many examples and train a set of fixing rules from them in advance that covers many common problems end users meet.
Another benefit is that the manual efforts of adapting the searched correction to their cases, which might be the most challenging part to end users, are saved.

\section{Rules in EqFix}\label{sec:lang}

\EqFix is a rule-based system.
Rules are formally defined by a DSL as shown in \cref{fig:core-lang}.
Each \emph{rule} $\mathcal{R}$ is a pair $\pair{EP, \mathcal{T}}$, where 
$EP$ is an \emph{error pattern} describing the template of the error message and which problem-specific information needs to be extracted from that message, and
$\mathcal{T}$ is a \emph{transformer} specifying the required transformation on the erroneous equation to fix the equation repair problem.

\begin{figure}[t]
    \begin{align*}
        \text{(Fixing) rule}~\mathcal{R} &::= \pair{EP, \mathcal{T}} \\
        \text{Transformer}~\mathcal{T} &::= \{v_1 \mapsto \tau_1, \ldots, v_k \mapsto \tau_k\} \\
        \text{Error pattern}~EP &::= [M_1, \ldots, M_k] \\
        \text{Matcher}~M &::= s \mid v \\
        \text{Equation pattern}~P &::= [M_1, \ldots, M_k]
    \end{align*}
    \caption{Syntax of fixing rules.}
    \label{fig:core-lang}
\end{figure}

\subsection{Error Pattern}\label{sec:error-pattern}

An error message either comes from the \LaTeX{} compiler (\eg \#5 -- \#8 of \cref{tb:motivating}) or the user (\eg \#1 -- \#4).
For the latter case, we assume the error message starts with one or more predefined \emph{keywords} (as in \cref{tb:keywords}) that mention the error type and then followed by a substring of the erroneous equation which locates the error.
In either case, we represent the error message as a natural language sentence that can be split into a list of \emph{tokens} $[e_1, \ldots, e_k]$ by delimiters (whitespaces, commas, etc.).
\EqFix is unaware of the resource of the error messages and employs a unified \emph{error pattern} to match against them.
Users are allowed to customize their keywords because \EqFix regards them as normal tokens.

An \emph{error pattern} $EP = [EM_1, \ldots, EM_k]$ contains a list of \emph{matchers}, where each of them is either:
(1) a string matcher $s$ that only matches against $s$ itself, or
(2) a variable matcher $v$ that matches against any token and binds the matched token to $v$.
An error message $[e_1, \ldots, e_{k'}]$ matches $EP$ if
they have the same length ($k = k'$), and that
every token $e_i$ ($1\le i\le k$) matches the corresponding matcher $M_i$.
The matching result (if succeeds) is a mapping from variables to the bound string values $\{v_1\mapsto s_1, \ldots, v_k\mapsto s_k\}$.

\begin{example}\label{ex:err-pattern}
    Consider $err_2 = [``\verb|superscript|", ``\verb|123|"]$ from \cref{tb:motivating}.
    Let $EP_1 \defas [``\verb|superscript|", v_1]$ be an error pattern.
    Matching $err_2$ against $EP_1$ succeeds with $\{v_1 \mapsto ``\verb|123|"\}$, \ie
    $``\verb|123|"$ is bound to $v_1$.
\end{example}

\subsection{Equation Pattern}\label{sec:lang:core:pattern}

Unlike an error message, an equation text usually involves complicated syntax and thus cannot be directly tokenized by commonly seen delimiters.
To extract the problem-specific information from an equation (\eg to find the cause of the error),
we propose the notion of an \emph{equation pattern}. 

An \emph{equation pattern} $P = [M_1, \cdots, M_k]$ consists of a list of matchers.
Especially, the string and variable matchers in $P$ must appear \emph{alternately},
that is, if $M_i$ is a string, then $M_{i+1}$ must be a variable and vice versa.
Intuitively, the string matchers in an equation pattern are indeed used as
``delimiters'' to tokenize the equation into a list of ``tokens''. 
Each ``token'' of the equation matches a variable matcher and may convey some
\emph{problem-specific} information which may be useful later in generating a corrected equation.
Oppositely, if we allow consecutive variable matchers to appear in an equation pattern, the split would be \emph{ambiguous}.
For instance, consider a string ``\verb|alpha|'' and an equation pattern $[v_1,v_2]$,
we could either let $v_1$ match against ``\verb|a|'' (and $v_2$ match against ``\verb|lpha|''),
or let $v_1$ match against ``\verb|al|'' (and $v_2$ match against ``\verb|pha|''), etc.

\paragraph*{Pattern Matching}

Given that string and variable matchers appear alternately, we pattern match an equation pattern $P$ against an equation $eq$ by simply locating the occurrences of the string matchers in $eq$ (failure if we cannot)---the variable matchers then match against the substrings in between.
For example, matching $[v_1, ``\verb|foo|", v_2]$ against ``\verb|(foo)|'' yields the bindings
$\{v_1 \mapsto ``\verb|(|", v_2 \mapsto ``\verb|)|"\}$,
because the equation is split into three parts $``\verb|(|" \concat ``\verb|foo|" \concat ``\verb|)|"$.

\paragraph*{Pattern Instantiation}

Equation patterns can be regarded as ``templates'' of equation text where the variable matchers are ``placeholders''.
Thus it is natural to define pattern \emph{instantiation}---the reverse of pattern matching---to obtain a (concrete) equation by replacing the variable matchers with the bound strings.

\begin{example}\label{ex:matcher}
    Consider example \#2 of \cref{tb:motivating}, where \( eq_2=``\verb|$y^123+x$|" \).
    Let \( P_2 \defas [``\verb|$y^|", v_1, ``\verb|+x$|"] \) be an equation pattern.
    Matching $eq_2$ against $P_2$ gives $\sigma=\{v_1 \mapsto ``\verb|123|"\}$.
    Further, instantiating $P_2$ with $\sigma$ gives back $eq_2$.
\end{example}

\paragraph*{Pattern Generation}

In \EqFix, equation patterns are not explicitly presented in the rule.
They are only \emph{intermediate} during rule application.
To construct an equation pattern $P$ from an erroneous equation $eq$ with the matching result $\sigma=\{v_1 \mapsto s_1, \ldots, v_k \mapsto s_k\}$ from an error pattern, 
we substitute all the occurrences of $s_1, \ldots, s_k$ in $eq$ with $v_1, \ldots, v_k$, respectively.

\begin{example}\label{ex:pattern-gen}
    Given $\sigma=\{v_1 \mapsto ``\verb|123|"\}$ and \( eq_2=``\verb|$y^123+x$|" \),
    applying the above process yields
    \( [``\verb|$y^|", v_1, ``\verb|+x$|"] \).
\end{example}

\subsection{Transformer}\label{sec:transformer}

Our DSL achieves an underlying repairing strategy via string transformation.
Since string matchers express the \emph{problem-unspecific} information, the substrings in the erroneous equation matched by them should be kept in the corrected equation.
The substrings matched by variable matchers, on the other hand, need to be transformed by \emph{string transformers}---functions that map a string into a new one.
\ifExt
Technical details of the string transformers will be presented in \cref{sec:transformer:string-trans}.
\else
The string transformer we employ in \EqFix is expressed by a variant of \FlashFill's DSL \cite{flash-fill} but possesses a more restricted syntax for better efficiency.
Technical details can be found in our extended version.
\fi
In \EqFix, we define a \emph{transformer}---a mapping from the variable matchers into string transformers---to collect all necessary string transformations.

\begin{example}\label{ex:transformer}
    Let $\tau_1$ be a string transformer that inserts a pair of curly parentheses surrounding the input.
    \ifExt
    Note: this string transformer will be defined in \cref{ex:st:1}.
    \fi
    Let $\mathcal{T}_1 \defas\{v_1 \mapsto \tau_1\}$ be a transformer.
    Applying $\mathcal{T}_1$ to $\sigma=\{v_1 \mapsto ``\verb|123|"\}$ yields
    $\sigma'=\{v_1 \mapsto ``\verb|{123}|"\}$.
\end{example}

\ifExt

\subsection{String Transformer}\label{sec:transformer:string-trans}

We now present $\mathcal{L}_\text{ST}$, a DSL for expressing the underlying string transformation.
We first define notions for manipulating substrings.
Let $s$ be a string, we denote $s[k_1..k_2]$ the substring of $s$ that
starts from the index $k_1$ (inclusive) until the index $k_2$ (inclusive).
We assume the indexes start from $0$, and we use a negative index $-k (k>0)$ to count from the end of the string.
For simplicity, we write $s[..k]$ for $s[0..k]$, and $s[k..]$ for $s[k..-1]$.

\begin{figure}[t]
    \begin{align*}
        \tau &::= \t{fun}~x \Rightarrow S \\
        S &::= F \mid \tk{Concat}(F, S) \\
        F &::= \tk{ConstStr}(s) \mid \tk{SubStr}(x, p_1, p_2) \\
        p &::= \tk{AbsPos}(x, k) \mid \tk{RelPos}(x, t, j, k)
    \end{align*}
    \caption{Syntax of the string transformer language $\mathcal{L}_\text{ST}$.}
    \label{fig:st-lang}
\end{figure}

The syntax of $\mathcal{L}_\text{ST}$ is described in \cref{fig:st-lang}.
A string transformer is a function that takes a string $x$ as input and produces a new string as output.
The function body is defined by a string expression $S$ that evaluates to a string.
A string expression is either an atomic expression $F$ in which no concatenation is allowed
or a concatenation of an atomic expression with another string expression.
An atomic expression $F$ is either of the form $\tk{ConstStr}(s)$ denoting a constant string $s$,
or $\tk{SubStr}(x, p_1, p_2)$ denoting the substring $x[i_1..i_2]$, given that the two position expressions $p_1$ and $p_2$ are evaluated to indexes $i_1$ and $i_2$ respectively.

A position expression $p$ evaluates to an index of a string $x$.
It is either of the form $\tk{AbsPos}(x, k)$ denoting the absolute index $k$,
or $\tk{RelPos}(x, t, j, k)$ denoting a relative index of the $j$-th occurrence of the regular expression token $t$ in $x$ plus an offset $k$.
We allow $j$ to be a negative value, say $-i(i>0)$, indicating the last $i$-th occurrence.
The same case also applies to the offset $k$.
A regular expression token can be customized to employ the domain knowledge of mathematical equations,
such as commonly-seen special characters
$``\verb|{|"$, $``\verb|}|"$, $``\verb|(|"$, $``\verb|)|"$, $``\verb|^|"$, $``\verb|_|"$ and numbers.

The following example illustrates the semantics of position expressions.
\begin{example}
    Let $x$ be bound to the string $``\verb|$t_{k_{i}^j}$|"$,
    \begin{itemize}
        \item $\tk{AbsPos}(x, -2)$ evaluates to 11 ($``\verb|}|"$);
        \item $\tk{RelPos}(x, ``\verb|_|", -1, 0)$ evaluates to 5 (last $``\verb|_|"$);
        \item $\tk{RelPos}(x, ``\verb|_|", -1, 1)$ evaluates to 6 (last
            $``\verb|_|"$ plus an offset 1);
        \item $\tk{RelPos}(x, ``\verb|_|", -1, -1)$ evaluates to 4 (last
            $``\verb|_|"$ plus an offset -1).
    \end{itemize}
\end{example}

\begin{example}\label{ex:st:1}
    The following gives the formal definition of a string transformer that inserts a pair of curly parentheses surrounding the input $x$:
    $$\tau_1=\t{fun}~x \Rightarrow \tk{Concat}(\tk{ConstStr}(\verb|{|),\tk{Concat}(x, \tk{ConstStr}(\verb|}|)))$$
    such that $\tau_1(``\verb|10|") = ``\verb|{10}|"$, and $\tau_1(``\verb|123|") = ``\verb|{123}|"$.
\end{example}

\paragraph{Related DSLs}

Our string transformation DSL $\mathcal{L}_{\text{st}}$ is inspired by \FlashFill and possesses a more restricted syntax:
it only allows a regular token to appear in the relative position,
whereas in \FlashFill, several regular expressions are allowed.
Apparently, $\mathcal{L}_{\text{st}}$ is less expressive than \FlashFill.
However, it is sufficient for expressing the string
transformations for equation repairing problems, where
the error parts have already been located by matchers.

Another simplified example of \FlashFill is \NoFAQ, which does not allow the concatenation of two substring expressions.
However, we find that string concatenation is a useful and necessary operation for
equation repair, and thus keep this operation in $\mathcal{L}_{\text{st}}$.
To sum up, $\mathcal{L}_{\text{st}}$ can be considered as an adaption of the \FlashFill's DSL to the
equation repair problem:
it is restricted for efficient searching, but
remains expressive for the equation repair problem.

\fi

\subsection{Rule Application}

Putting the above operations together, we now present how a rule $\mathcal{R} = \pair{EP, \mathcal{T}}$ is applied to an equation repair problem $(eq, err)$:
\begin{enumerate}
    \item match $EP$ against $err$ to extract problem-specific information recorded in a mapping $\sigma$;
    \item generate an equation pattern $P$ (from $\sigma$), regarded as an ``template'' of $eq$;
    \item perform the underlying repairing strategy expressed by the set of string transformers in $\mathcal{T}$ on $\sigma$ to obtain a new $\sigma'$;
    \item obtain the corrected equation by instantiating $P$ with $\sigma'$.
\end{enumerate}

\begin{figure}[t]
    \centering
    \tikzstyle{line} = [draw, -latex']
\tikzstyle{proc} = [rectangle, draw, text width=6em, text centered]

\begin{tikzpicture}[node distance=5ex and 7em, every path/.style={->}]
    \node (err) {$
        \begin{aligned}
            err_2 &:[``\verb|superscript|", ``\verb|123|"] \\
            EP_1&:[``\verb|superscript|", v_1]
        \end{aligned}
    $};
    \node (env1) [below=of err] {$\{v_1 \mapsto ``\verb|123|"\}$};
    \node (trans) [proc, text width=1em, below=of env1] {$\mathcal{T}_1$};
    \node (gen) [proc, right=of env1] {Pattern Generation};
    \node (eq) [above=of gen]  {\( eq_2:``\verb|$y^123+x$|" \)};
    \node (pat) [below=of gen] {\( P_2:[``\verb|$y^|", v_1, ``\verb|+x$|"] \)};
    \node (env2) [below left=6ex and 4.8em of pat] {$\{v_1 \mapsto ``\verb|{123}|"\}$};
    \node (ins) [proc, below=of pat] {Pattern Instantiation};
    \node (fix) [right=4em of ins] {\( fix_2:``\verb|$y^{123}+x$|" \)};

    \path [line] (err) -- node [right] {\textsf{Step 1}} (env1);
    \path [line] (eq) -- (gen);
    \path [line] (env1) -- (gen);
    \path [line] (gen) -- node [right] {\textsf{Step 2}} (pat);
    \path [line] (env1) -- (trans);
    \path [line] (trans) -- node [right] {\textsf{Step 3}} (env2);
    \path [line] (pat) -- (ins);
    \path [line] (env2) -- (ins);
    \path [line] (ins) -- node [above] {\textsf{Step 4}} (fix);
\end{tikzpicture}
    \caption{Application of the rule $\mathcal{R}_1=\pair{EP_1,\mathcal{T}_1}$ to the input $(eq_2,err_2)$.}
    \label{fig:apply}
\end{figure}
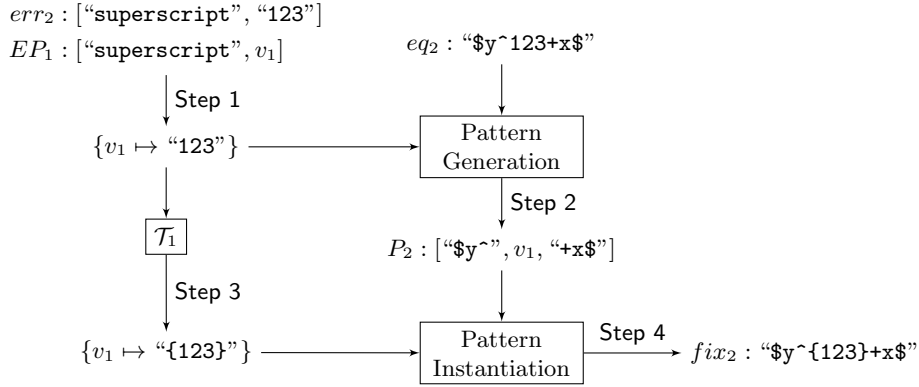

\begin{example}
    Let rule $\mathcal{R}_1 \defas \pair{EP_1, \mathcal{T}_1}$,
    where $EP_1$ is defined in \cref{ex:err-pattern} 
    and $\mathcal{T}_1$ is defined in \cref{ex:transformer}.
    Following the above steps, we apply $\mathcal{R}_1$ to the equation repair problem $(eq_2, err_2)$ (depicted by \cref{fig:apply}):
    \begin{enumerate}
        \item matching $EP_1$ against $err_2$ gives $\sigma$ (as in \cref{ex:err-pattern});
        \item generate $P_2$ from $eq_2$ and $\sigma$ (as in \cref{ex:pattern-gen});
        \item transform $\sigma$ into $\sigma'$ by $\mathcal{T}_1$ (as in \cref{ex:transformer});
        \item instantiating $P$ with $\sigma'$ gives $fix_2$ (as in \cref{ex:matcher}).
    \end{enumerate}
\end{example}

\section{Rule Synthesis}\label{sec:syn}

The rule synthesis algorithm takes a set of input-output examples $\mathcal{E}$ as the specification,
and generates a fixing rule $\mathcal{R} = \pair{EP, \mathcal{T}}$ consistent with the examples.
The synthesis consists of two passes:
(1) an error pattern $EP$ is synthesized from the examples $\mathcal{E}$, and 
(2) a transformer $\mathcal{T}$ is synthesized from $\mathcal{E}$ and $EP$.

\subsection{Synthesizing Error Patterns}\label{sec:syn:err-pattern}

Given an error message $err=[e_1, \ldots, e_k]$, our problem is to generate an error pattern $EP$ (of the same length) that matches against $err$.
To achieve this goal, a naive pattern $EP_{\bot} = [e_1, \ldots, e_k]$, the error message itself, is apparently a solution.
However, it is so \emph{restricted} that only this error message can match it.
Oppositely, another naive pattern $EP_{\top} = [v_1, \ldots, v_k]$ is too \emph{general} and can be matched with any error message with $k$ tokens.

To synthesize an error pattern that is neither too restricted nor too general,
we start with $EP_{\bot}$, and for each string in $EP_{\bot}$, replace it with a fresh variable;
if it also occurs in either the input equation $eq$ or the output equation $fix$.
Such a replacement makes the error pattern more general.
Realizing that the error message is usually related to either the input equation by telling why it is erroneous or the output equation by explaining how to repair it, the introduced variables, in either case, will capture such important information.

\ifExt

\begin{example}\label{ex:syn:err-pattern-1}
    Consider example \#1 of \cref{tb:motivating},
    where the error message is
    $$err_1=[``\verb|superscript|", ``\verb|10|"].$$
    Comparing $err_1$ to $eq_1$ and $fix_1$,
    apparently only the token $``\verb|10|"$ is a common substring.
    Thus we synthesize an error pattern:
    $EP_1=[``\verb|superscript|", v_1]$.
\end{example}

\fi

\subsection{Synthesizing Transformers}\label{sec:syn:transformer}

The essential problem of synthesizing a transformer $\{v_1 \mapsto \tau_1, \ldots, v_k \mapsto \tau_k\}$ is to synthesize the underlying string transformers $\tau_1, \ldots, \tau_k$.
This problem has been well-studied in previous literature, and a well-known approach could be the PBE approach invented by \FlashFill \cite{flash-fill}.
To adopt their approach in our setting, we must extract a set of \emph{input-output string examples},
each is a pair $(s, s')$ packed the input string $s$ and the expected output string $s'$,
as the specification for synthesizing each string transformer $\tau_i$.
\ifExt
Our synthesis algorithm for $\mathcal{L}_\text{ST}$ is also an adaption of \FlashFill's, which will be presented in \cref{sec:syn:st}.
For now, we focus on how to extract the input-output string examples.
\fi

Let us first consider the situation where the generated pattern (using the pattern generalization process mentioned in \cref{sec:lang:core:pattern}) matches against both the input and output equations for all examples $\mathcal{E}$.
This condition implies that we can always compute two mappings by matching the generated pattern against the input and output equation.
Thus, to synthesize $\tau_i$, we are able to extract an input-output string example $(\sigma(v_i), \sigma'(v_i))$ for each example in $\mathcal{E}$ (so the complete specification is their union), where $\sigma$ and $\sigma'$ are the two mappings computed as above.

\begin{example}\label{ex:syn:st-examples}
    Given examples \#1 and \#2 from \cref{tb:motivating}, to synthesize a transformer $\tau_1$, 
    we extract $(``\verb|10|", ``\verb|{10}|)$ from example \#1, and
    $(``\verb|123|", ``\verb|{123}|")$ from example \#2 (see \cref{ex:pattern-gen} for the generated pattern).
    Thus, the complete specification for synthesizing $\tau_1$ is
    $\varphi_1 = \{ (``\verb|10|", ``\verb|{10}|"), (``\verb|123|", ``\verb|{123}|") \}$.
\end{example}

Suppose the generated pattern only matches against the input equation but not the output; we must generalize this pattern so that it matches against the output equation.
The generalization process is an extension to \EqFix, and we will discuss it later in \cref{sec:syn:relax}.
Once this is done, the synthesis method we have just introduced works again.
So far, we have adequate mechanism to synthesize a rule from example \#1 of \cref{tb:motivating} (depicted in \cref{fig:synthesis}):

\begin{figure}[t]
    \centering
    \tikzstyle{line} = [draw, -latex']
\tikzstyle{proc} = [rectangle, draw, text width=10em, text centered]

\begin{tikzpicture}[node distance=4ex and 5em, every path/.style={->}]
    \node (ep-syn) [proc, text width=11em] {Error Pattern Synthesis};
    \node (example) [left=2em of ep-syn] {$(eq_1,err_1,fix_1)$};
    \node (ep) [right=4em of ep-syn] {$EP_1$};
    \node (env1) [below=of ep-syn] {$\{v_1 \mapsto ``\verb|10|"\}$};
    \node (gen) [proc, text width=9em, left=of env1] {Pattern Generation};
    \node (pat) [below=of gen] {\( P_1:[``\verb|$x^|", v_1, ``\verb|$|"] \)};
    \node (match) [proc, text width=9em, below=of pat] {Pattern Matching};
    \node (t-syn) [proc, below=of env1] {Transformer Synthesis};
    \node (env2) [below=of t-syn] {$\{v_1 \mapsto ``\verb|{10}|"\}$};
    \node (t) [right=of t-syn] {$\mathcal{T}_1$};

    \path [line] (example) -- (ep-syn);
    \path [line] (ep-syn) -- node [above] {\textsf{Step 1}} (ep);
    \path [line] (ep-syn) -- (env1);
    \path [line] (env1) -- (gen);
    \path [line] (gen) -- node [right] {\textsf{Step 2}} (pat);
    \path [line] (pat) -- (match);
    \path [line] (match) -- node [below] {\textsf{Step 3}} (env2);
    \path [line] (env1.south) -- (t-syn);
    \path [line] (env2.north) -- (t-syn);
    \path [line] (t-syn) -- node [above] {\textsf{Step 4}} (t);
\end{tikzpicture}
    \caption{Synthesis a rule $\mathcal{R}_1=\pair{EP_1,\mathcal{T}_1}$ by example $(eq_1,err_1,fix_1)$.}
    \label{fig:synthesis}
\end{figure}
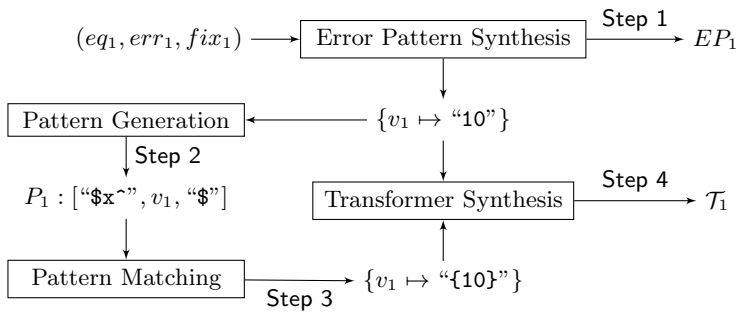

\begin{enumerate}
    \item synthesize an error pattern $EP_1$ by comparing $err_1$ with $eq_1$
        and $fix_1$, respectively, 
        $\sigma_1=\{v_1 \mapsto ``\verb|10|"\}$ records the values of the matched variable;
    \item generate the equation pattern $P_1$;
    \item match $P_1$ against $fix_1$, yielding $\sigma_1'=\{v_1 \mapsto ``\verb|{10}|"\}$;
    \item synthesize $\tau_1$ from $\{(``\verb|10|", ``\verb|{10}|")\}$, which gives rise to the transformer $\mathcal{T}_1 = \{v_1 \mapsto \tau_1\}$.
\end{enumerate}

\ifExt

\subsection{Synthesizing String Transformers}\label{sec:syn:st}

Before presenting how we synthesize string transformers, we need some background knowledge on version space algebra and the synthesis method based on this kind of program representation.

\subsubsection{Preliminary: VSA}

\emph{Version space algebra} (VSA) is a succinct way to represent a large number of programs in polynomial space.
Such a succinct representation is demanding for our synthesis problem because there can be a large set of candidate string transformers consistent with the specification.
The VSA approach we adopt here has been well-studied in the recent decade \cite{flash-fill,flash-meta,flash-extract}.

Intuitively, a VSA can be viewed as a directed graph, in which every node represents a set of programs.
A \emph{leaf node} is explicitly annotated with a set of concrete programs, whereas an internal node uses an implicit representation.
Two typical kinds of internal nodes are:
(1) \emph{union nodes} for the set-theoretic union of the program sets represented by its children;
(2) \emph{join nodes} for Cartesian-product over all possible applications of a constructor over its arguments, each individually chosen from the program set which the child node represents.
We denote the set of concrete programs represented by a VSA $\tilde{N}$ as $\sem{\tilde{N}}$.
A program $P$ is included in a VSA $\tilde{N}$ iff $P \in \sem{\tilde{N}}$.

When a set of programs are synthesized and succinctly represented by a VSA, to find the desired ones, it is impractical to enumerate the entire set of the candidate programs.
Thus, \emph{ranking functions} are designed for fast convergence.
With a proper ranking function, programs that meet the user's intent are very likely to be ranked at the top.

\subsubsection{VSA Representation of Rules}

We lift our rule DSL to the VSA representation in the following way:
\begin{align*}
    \tilde{\mathcal{R}} &::= \pair{EP, \tilde{\mathcal{T}}} \\
    \tilde{\mathcal{T}} &::= \{v_1 \mapsto \tilde{\tau_1}, \ldots, v_k \mapsto \tilde{\tau_k}\}
\end{align*}
where $\tilde{\tau}$, the VSA representation of $\mathcal{L}_\text{ST}$ is as follows:
$$\begin{array}{rcl}
    \tilde{\tau} &::=& \t{fun}~\gamma\Rightarrow\tilde{S} \\
    \tilde{S} &::=& \tilde{F} \mid \tk{Concat}_{\join}(\tilde{F}, \tilde{S}) \\
    \tilde{F} &::=& \tk{ConstStr}(s) \mid 
                    \Let~x=\gamma_i~\In~\tk{SubStr}_{\join}(x, \tilde{p_1}, \tilde{p_2})
\end{array}$$
where the position VSA $\tilde{p}$ uses explicit representation.

The set of concrete programs is defined as:
\begin{align*}
    \sem{\pair{EP, \tilde{\mathcal{T}}}} &=
        \{ \pair{EP, \mathcal{T}} \mid \mathcal{T} \in \sem{\tilde{\mathcal{T}}} \} \\
    \sem{\{v_1 \mapsto \tilde{\tau_1}, \ldots, v_k \mapsto \tilde{\tau_k}\}} &=
        \{\{v_1 \mapsto \tau_1, \ldots, v_k \mapsto \tau_k\} \mid \tau_1 \in \sem{\tilde{\tau_1}}, \ldots, \tau_k \in \sem{\tilde{\tau_k}} \} \\
    &\\
    \sem{\t{fun}~\gamma\Rightarrow\tilde{S}} &= \{ \t{fun}~\gamma\Rightarrow S \mid S \in \sem{\tilde{S}} \} \\
    \sem{\tk{Concat}_{\join}(\tilde{F}, \tilde{S})} &= 
        \{ \tk{Concat}(F, S) \mid F \in \sem{\tilde{F}}, S \in \sem{\tilde{S}} \} \\
    \sem{\tk{ConstStr}(s)} &= \{ \tk{ConstStr}(s) \} \\
    \sem{\tk{SubStr}_{\join}(x, \tilde{p_1}, \tilde{p_2})} &=
        \{ \tk{SubStr}(x, p_1, p_2) \mid p_1 \in \sem{\tilde{p_1}}, p_2 \in \sem{\tilde{p_2}} \}
\end{align*}
where $\tilde{\mathcal{T}}$ is a join node.

\begin{example}\label{ex:syn:transformer}
    Given the specification $\varphi_1$, 
    let $\tilde{\tau_1}$ be the VSA returned by the underlying synthesizer.
    Then, the transformer VSA is
    $\tilde{\mathcal{T}_1}=\{ v_1 \mapsto \tilde{\tau_1} \}$,
    and the rule VSA is
    $\tilde{\mathcal{R}_1}=\pair{EP_1, \tilde{\mathcal{T}_1}}$.
    Furthermore,
    we observe that the string transformer $\tau_1$ defined in \cref{ex:st:1} 
    is consistent with $\varphi_1$, and $\tau_1 \in \sem{\tilde{\tau_1}}$.
    Thus, the rule $\mathcal{R}_1=\pair{EP_1, \mathcal{T}_1} \in \sem{\tilde{\mathcal{R}_1}}$
    is consistent with the examples \#1 and \#2 in \cref{tb:motivating}.
\end{example}

\subsubsection{Synthesis}

At a high level, the string transformer synthesis algorithm takes a set of input-output string examples as input,
and generate a set of candidate string transformers represented by a VSA $\tilde{\tau}$ as output.
The basic idea of VSA-based synthesis is to divide the problem into several sub-problems, each for a nonterminal appearing in the DSL.
In our situation, we must propose algorithms to find the set of subprograms expressed by string expressions ($S$),
atomic expressions ($F$), and position expressions ($p$), respectively, that satisfy the input specification.
For the string expression case, the specification is simply the input of the entire synthesis algorithm.
For the other two cases, we need manually specify the so-called \emph{witness functions} to obtain the corresponding specification.

Like \FlashFill, a program expressed by $\mathcal{L}_\text{ST}$ with a ``simpler'' structure is preferred.
For instance, we give a string expression concatenated with fewer atomic expressions a higher rank than another one with more atomic expressions.
We also give a relative position expression a higher rank than an absolute position expression because the former is considered to be more general and fits more problems.

\cref{fig:st-syn-algo} describes our synthesis algorithm for
$\mathcal{L}_{\text{st}}$.
The entry function $\t{genString}$ takes the inputs $\gamma$ and an output string $s$
as the specification, and generates all string expressions 
that are consistent with the specification, 
in terms of either an atomic expression or a \tk{Concat},
say the string expression evaluates to $s$, given $\gamma$.
These expressions form a VSA and can be classified into two groups:
(1) an atomic expression that evaluates to $s$, which is computed by \t{genAtomic};
(2) an atomic expression $F$ which evaluates to $f$, 
concatenate another string expression $S$ which evaluates to $s'$,
such that $s = f \concat s'$.

Similarly, we interpret the other generation function as follows:
\begin{itemize}
    \item $\t{genAtomic}~\gamma~s$ generates all atomic expressions that evaluates to $s$
    given the input $\gamma$;
    \item $\t{genSubStr}~x~s$ generates all substring expressions that evaluates to $s$
    given the input string $x$;
    \item $\t{genPos}~x~k$ generates all position expressions that evaluates to $k$
    given the input string $x$.
\end{itemize}

When multiple string examples are provided, we could generate a VSA for each string example,
and then intersect all these VSAs to produce a VSA that is consistent with all the examples.

\begin{figure}[t]
    \begin{align*}
        &\t{genString}~(\gamma:\t{string list})~(s:\t{string}) \\
            &\quad= \t{genAtomic}~\gamma~s \\
            &\quad\cup \{\tk{Concat}_{\join}
            (\t{genAtomic}~\gamma~s[..i], \t{genString}~\gamma~s[i+1..]) \mid 1<i<\t{len}~s \} \\
        &\t{genAtomic}~(\gamma:\t{string list})~(s:\t{string}) \\
            &\quad= \{\tk{ConstStr}(s)\} \cup \{ \t{genSubStr}~x~s \mid x \in\gamma \} \\
        &\t{genSubStr}~(x:\t{string})~(s:\t{string}) \\
            &\quad= \{\tk{SubStr}_{\join}(x, \t{genPos}~x~k, \t{genPos}~x~(k+\t{len}~s-1)) \mid k \in K\} \\
            &\quad \text{where}~K = \{k \mid s = x[k..k+\t{len}~s-1]\} \\
        &\t{genPos}~(x:\t{string})~(k:\t{int}) \\
            &\quad= \{\tk{AbsPos}(x, k), \tk{AbsPos}(x, k - \t{len}~s)\} \cup \t{genRelPos}~x~k
    \end{align*}
    \caption{Synthesis algorithm for $\mathcal{L}_\text{ST}$.
    Function $\t{len}~s$ givens the length of string $s$.
    The implementation of $\t{genPos}~x~k$,
    which generates all position expressions in terms of \tk{RelPos}
    that evaluates to $k$ given the input string $x$,
    depends on the selection of regular tokens,
    and thus we leave it abstract here.}
    \label{fig:st-syn-algo}
\end{figure}

\fi

\subsection{Extension: Pattern Generalization via Lazy Relaxation}\label{sec:syn:relax}

Let us study some examples to get a sense of how to generalize patterns.

\begin{example}\label{case:relaxer:mot-1}
Consider example \#5 of \cref{tb:motivating}, where
\begin{itemize}
    \item \(eq_5 = ``\verb|${1,2,3$|"\),
    \item \(fix_5=``\verb|${1,2,3}$|"\), and
    \item \(err_5 = ``\verb|Missing } inserted|"\).
\end{itemize}
Let $EP_2 \defas [``\verb|Missing|", v_1, ``\verb|inserted|"]$ be the error pattern.
Matching $EP_2$ against $err_5$ gives $\sigma_5=\{ v_1 \mapsto ``\verb|}|" \}$.
From $eq_5$ and $\sigma_5$, an equation pattern \( P_5=[``\verb|${1,2,3$|"] \) is generated.
We see that $P_5$ cannot be matched by $fix_5$.
However, a more general pattern such as $[v]$ (where $v$ is a fresh variable) matches with $fix_5$.
\end{example}

\begin{example}\label{case:relaxer:mot-2}
Consider
    $eq = s_1 \concat s_2 \concat s_3 \concat s_4$, 
    $fix = s_1 \concat s_2' \concat s_3' \concat s_4'$, 
and $\sigma = \{ v_1 \mapsto s_2, v_2 \mapsto s_4 \}$.
Note that $s_2$, $s_3$ and $s_4$ are all modified in $fix$ compared to
$eq$, however $s_3$ is not bound to a variable in $\sigma$.
From $eq$ and $\sigma$, an equation pattern $[s_1, v_1, s_3, v_2]$ is generated,
which apparently does not match against $fix$. 
However, a more general pattern like $[s_1, v]$ ($v$ is a fresh variable)
can match $fix$, where $v$ is matched against $s_2' \concat s_3' \concat s_4'$.
\end{example}

We learn from the examples that if the generated equation pattern cannot match the output equation,
we can always replace several string matchers with fresh variables until it matches against the output equation -- we call this process \emph{pattern relaxation}.
As the relaxation goes on, the pattern becomes more and more general.
In the worst case, it gives $P_\top=[v]$ that consists of a single variable $v$ and can be matched by any string (as in \cref{case:relaxer:mot-1}).
To find a relaxed pattern that is as strict as possible, the relaxation should be \emph{lazy}.

\paragraph*{Lazy Relaxation Process}

Let $P$ be an equation pattern, and $s$ be a string (\ie the output equation).
As a special case, if $P$ is a constant (\ie contains no variables), $P_\top$ is the only possible relaxed pattern.
Otherwise, since $P$ cannot match against $s$, there must be some string $s'$ in $P$
that is not a substring of $s$.
Three kinds of relaxations are performed on $P$ depending on the relative position (left, right, or binary) of $s'$ in $P$:
\begin{enumerate}[]
    \item (left) If $s'$ appears at the beginning of $P$,
    we replace the subpattern that consists of $s'$ and the variable $V$ followed by it,
    with a fresh variable $\tk{LVar}(V)$.
    \item (right) If $s'$ appears in the end of $P$,
    we replace the subpattern that consists of $s'$ and the variable $V$ before it,
    with a fresh variable $\tk{RVar}(V)$.
    \item (binary) If $s'$ appears in the middle of $P$,
    we replace the subpattern that consists of $s'$ and the adjacent variables $V_1$ and $V_2$,
    with a fresh variable $\tk{BVar}(V_1,V_2)$.
\end{enumerate}
The above repeats until the current pattern already matches against $s$.

\begin{example}\label{ex:relax:special}
    Pattern $P_5$ in \cref{case:relaxer:mot-1} is relaxed to $P_\top$. 
\end{example}
\begin{example}\label{ex:relax:easy}
    Pattern $[s_1$, $v_1, s_3$, $v_2]$ in \cref{case:relaxer:mot-2} is relaxed to $[s_1, \tk{BVar}(v_1, v_2)]$.
\end{example}

\begin{example}\label{ex:relax:hard}
    Consider a pattern
    $[s_1, v_2, s_3, v_4, s_5, v_6, s_7]$
    and an output equation 
    $s_1' \concat s_2 \concat s_3 \concat s_4 \concat s_5' \concat s_6 \concat s_7'$.
    It takes several steps to obtain a relaxed result.
    In the following, we highlight the relaxed subpattern with an underline and annotate the unmatched string with an asterisk:
    \begin{align*}
        [\underline{s_1^{*}, v_2}, s_3, v_4, s_5, v_6, s_7]
        &\to [\tk{LVar}(v_2), s_3, \underline{v_4, s_5^{*}, v_6}, s_7] \\
        &\to [\tk{LVar}(v_2), s_3, \underline{\tk{BVar}(v_4, v_6), s_7^{*}}] \\
        &\to [\tk{LVar}(v_2), s_3, \tk{RVar}(\tk{BVar}(v_4, v_6))]
    \end{align*}
\end{example}

\ifExt
\paragraph*{DSL Extension}
\else
\paragraph*{Relaxers and Synthesis}
\fi

We extend the rule DSL (\cref{fig:core-lang}) to include \emph{relaxers} that syntactically encode the three kinds of relaxations, with \tk{id} for no relaxation:
\begin{align*}
    \text{Rule}~\mathcal{R} &::= \pair{EP, \{r_1, \ldots, r_k\}, \mathcal{T}} \\
    \text{Relaxer}~r &::= \tk{id}(v) \mid \tk{LRelax}(r) \mid \tk{RRelax}(r) \mid 
        \tk{BRelax}(r_1, r_2) \\
    \text{Variable}~V &::= v \mid \tk{LVar}(V) \mid \tk{RVar}(V) \mid \tk{BVar}(V_1, V_2)
\end{align*}
In applying a rule with relaxers, one or more subpatterns of the generated equation pattern will be substituted with fresh variables according to the relaxers.
\ifExt

\paragraph*{Relaxer Synthesis}
To synthesize a rule with relaxers,
we need an extra step that generates the relaxers after we have synthesized the equation pattern $P$ for some input-output example $(eq, err, fix)$.
Suppose that $P$ already matches against $fix$,
then no relaxation is needed and an empty set of relaxers is returned.
Otherwise, we follow the lazy relaxation process to generate a relaxed pattern $\hat{P}$.
Note that in $P$, \tk{LVar}, \tk{RVar} and \tk{BVar} variables can be used to trace the relaxation process,
we can ``reproduce'' the relaxing process by directly translating them into relaxers:

\begin{itemize}
    \item For \cref{ex:relax:special}, $r_\top$ is the synthesized relaxer.
    \item For \cref{ex:relax:easy}, $\tk{BRelax}(\tk{id}(v_1), \tk{id}(v_2))$ is synthesized from $\tk{BVar}(v_1, v_2)$.
    \item For \cref{ex:relax:hard}, the synthesized relaxers are $\{r_1, r_2\}$, where $r_1=\tk{LRelax}(\tk{id}(v_2))$ is translated from $\tk{LVar}(v_2)$,
        and $r_2=\tk{RRelax}(\tk{BRelax}(\tk{id}(v_4), \tk{id}(v_6)))$ is
        from $\tk{RVar}(\tk{BVar}(v_4, v_6))$.
\end{itemize}

Suppose multiple examples are provided, the synthesized relaxers will be updated in each iteration.
In the first iteration, the first example is processed as above, yielding a set of relaxers $R_1$.
When processing the $i$-th ($i \ge 2$) example $(eq_i,err_i,fix_i)$,
the previously synthesized relaxers $R_{i-1}$ are inherited and applied on the equation pattern $P_i$, generating $P_i'$.
The relaxation is then performed on $P_i'$ instead of $P_i$.
The relaxed pattern $\hat{P_i}$ is then translated into $R_i$.
The algorithm terminates when all examples have been processed.

\else
For synthesis, the above lazy relaxation process is performed, and the relaxations that have been applied are recorded as corresponding relaxers.
For more technical details, please refer to our extended version.
\fi
\section{Evaluation}\label{sec:eval}

We prototyped the proposed approach as a tool \EqFix, written in a combination of F\# and C\#, running on the .NET core platform.
Rule application, synthesis algorithms, and relaxer extensions were built following the approaches proposed in the paper.
We developed the synthesizer for string transformation under the PROSE framework (proposed in \cite{flash-meta}), in which we specified the syntax and semantics of our DSL, a set of witness functions for guiding PROSE's synthesis engine, and a bunch of scoring functions for ranking candidate programs.

To measure the performance of \EqFix,
we conducted an experimental comparison with \FlashFill \cite{flash-fill}, a state-of-the-art synthesizer for string manipulation,
on a dataset that consists of 89 input-output \emph{example groups} collected from the web (online help forums, tutorials, and technical blogs),
each reveals one type of common mistake that users make, such as mismatch of delimiters and misuse of commands.
Unlike the machine learning approaches, PBE techniques usually only need a few (2 -- 5) examples.
The lengths of the erroneous equations vary from 5 to 166, with an average of 18.

\subsection{Experimental Setup}

The baseline tool \FlashFill was initially designed for string manipulation in spreadsheets (such as Excel), so an input-output example comprises a column of strings as input and a single string as output.
To adapt \FlashFill to our problem domain, we regarded an erroneous equation and an error message as two \emph{indistinguished} input columns.

Another difference between \FlashFill and \EqFix is that \FlashFill does not maintain a rule library.
To make a fair comparison, we made the following adaption to avoid the usage of rule libraries:
both tools were tested on an equation repair problem immediately after the rules were synthesized using the examples under the same example group.
Since the number of consistent string transformers with a given specification is usually multiple, our synthesis algorithm will produce multiple candidate rules as well for one group of input-output examples.
The candidate rules were ranked with heuristics, and we only attempted the top-10 candidate rules for each test case for fixing.
If any rule gives a fixed equation that equals the expected correction of that test case, the test case is said ``solved'' (otherwise ``failed'').
In each example group, we left the one with the longest erroneous equation as the test case and the others as training examples.
We set four training configurations C1, \ldots, C4, where C$i$ ($i=1,\ldots,4$) means the first $i$ shortest (by the length of the erroneous equation) examples in the training set are used for synthesis.
There were, in total, 356 training runs of \EqFix (and also for \FlashFill).

The experiments were conducted on an Intel(R) Core(TM) i5 laptop with \SI{2.3}{GHz} CPU and \SI{8}{GB} memory, running Mac OS 11.6 and .NET core 2.2.207.

\subsection{Results}

\begin{figure*}[t]
    \begin{minipage}[t]{.47\textwidth}
        \centering
        \includegraphics[width=\textwidth]{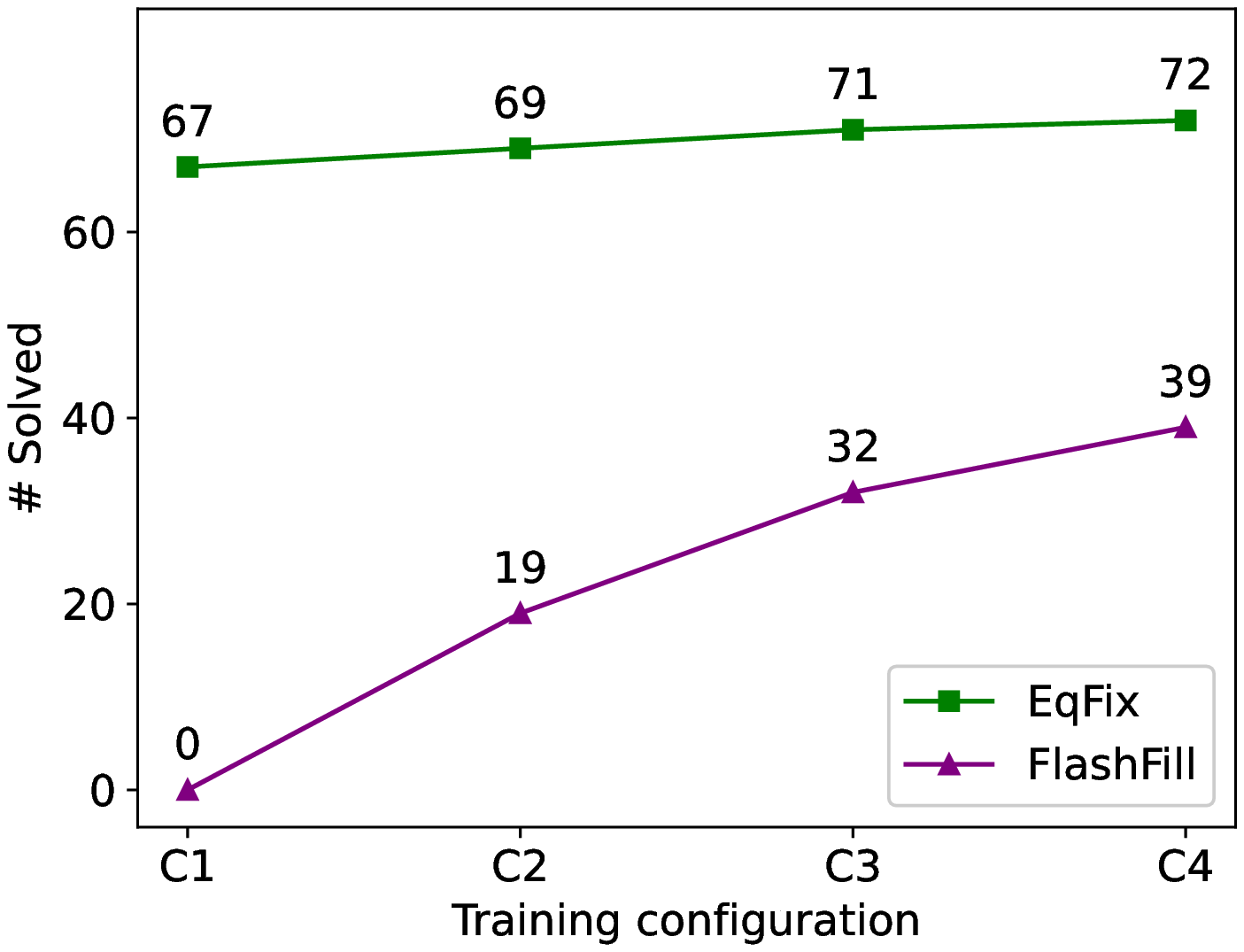}
        \caption{Overall number of solved test cases.}
        \label{fig:solved}
    \end{minipage}
    \hfill
    \begin{minipage}[t]{.47\textwidth}
        \centering
        \includegraphics[width=\textwidth]{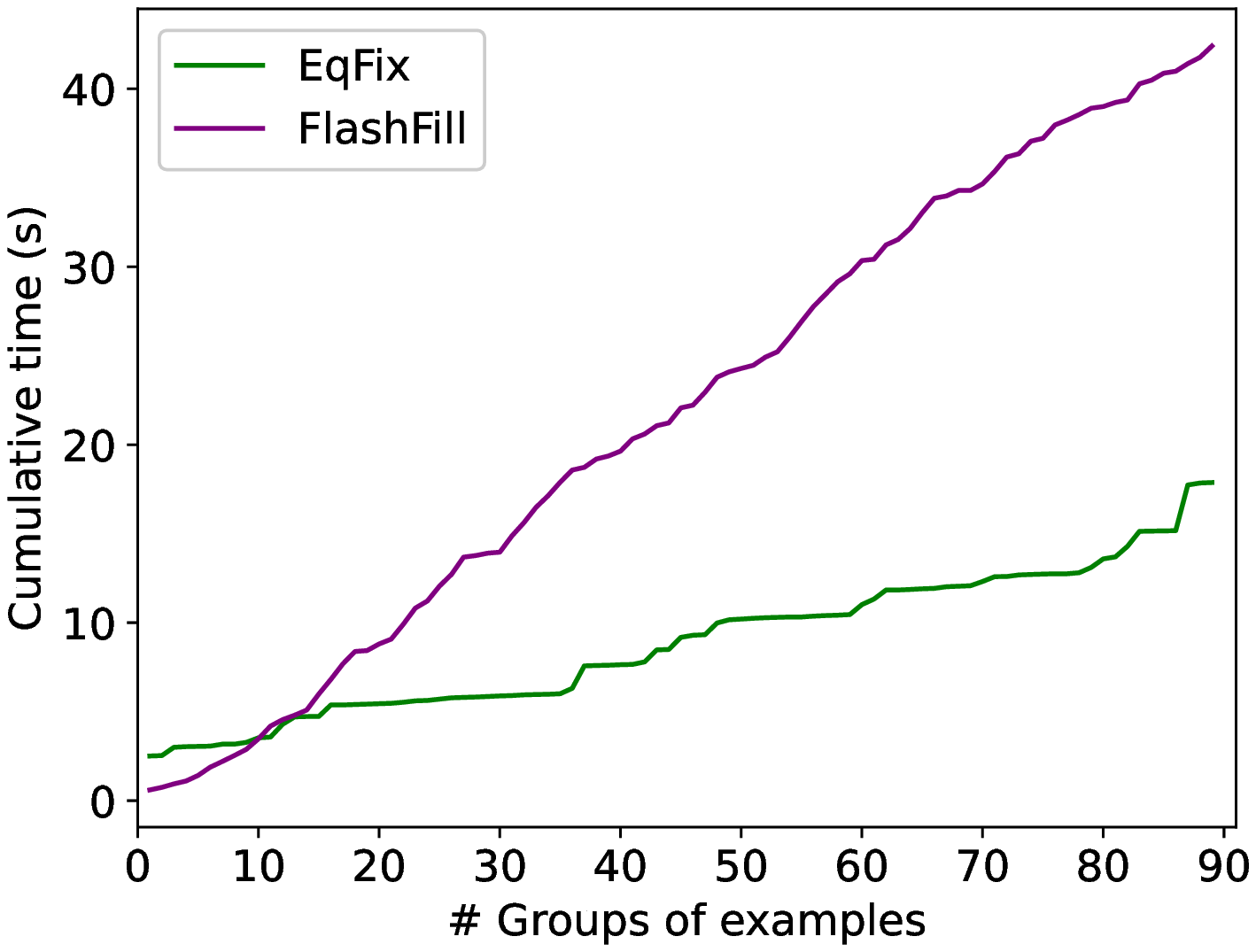}
        \caption{Cumulative synthesis time (s) with an increasing number of training example groups (C4).}
        \label{fig:syn-time}
    \end{minipage}
\end{figure*}

The overall number of solved test cases is presented in \cref{fig:solved}.
\EqFix outperformed \FlashFill under all configurations.
When training with only one example (C1), \EqFix solved 67 (75.2\%) test cases, whereas \FlashFill failed to solve any.
Both \EqFix and \FlashFill performed better when more training examples were given---this is well-understood, as more training examples eliminate spurious rules and lead to more general rules.
With the entire training set (C4), \EqFix solved 72 (80.9\%) while \FlashFill solved less than half of the test cases.
Therefore, \EqFix has a more vital learning ability in our problem domain.

\paragraph*{Impact of Ranking}

To understand how ranking affects the ability to solve the test cases, we list the numbers of attempted rules\footnote{Rules were attempted in the order of the rank list.} in \cref{tb:exp1}.
A cross mark ``\Failed'' indicates that no (top-10) rules produced the expected fix.
Among the solved test cases by \EqFix, at most 5 rules were attempted (\#81), and 57 -- 68 test cases were solved by the top-ranked rule.
In contrast, \FlashFill attempted at most 8 rules (\#48), and 0 -- 33 test cases were solved by the top-ranked rule.
We also recognize that providing more training examples helps to decrease the number of attempts, \eg in \#4 and \#82.

\begin{table*}[tbp]
    \caption{Numbers of attempted rules for solving each test case, \EqFix (E) v.s. \FlashFill (F).}
    \label{tb:exp1}
    \begin{center}
    \small

\begin{tabular}{c|cccccccc||c|cccccccc||c|cccccccc}
\toprule
\multirow{2}{*}{\#}     & \multicolumn{2}{c}{C1} & \multicolumn{2}{c}{C2} & \multicolumn{2}{c}{C3} & \multicolumn{2}{c||}{C4}                 & \multirow{2}{*}{\#}     & \multicolumn{2}{c}{C1}           & \multicolumn{2}{c}{C2} & \multicolumn{2}{c}{C3} & \multicolumn{2}{c||}{C4} & \multirow{2}{*}{\#} & \multicolumn{2}{c}{C1} & \multicolumn{2}{c}{C2} & \multicolumn{2}{c}{C3} & \multicolumn{2}{c}{C4} \\ \cmidrule(lr){2-9} \cmidrule(lr){11-18} \cmidrule(l){20-27} 
                        & E          & F         & E          & F         & E          & F         & E       & F       &                         & E & F       & E          & F         & E          & F         & E          & F         &                     & E          & F         & E          & F         & E          & F         & E          & F         \\ \midrule
1  & 1       & \Failed & 1       & \Failed & 1       & \Failed & 1       & \Failed & 31 & 1       & \Failed & 1       & 1       & 1       & 1       & 1       & 1       & 61 & 3       & \Failed & 3       & 2       & 3       & 2       & 3       & 1       \\
2  & 1       & \Failed & 1       & \Failed & 1       & \Failed & 1       & \Failed & 32 & 2       & \Failed & 1       & \Failed & 1       & \Failed & 1       & \Failed & 62 & 1       & \Failed & 1       & 1       & 1       & 1       & 1       & 1       \\
3  & \Failed & \Failed & \Failed & \Failed & \Failed & \Failed & \Failed & 1       & 33 & 2       & \Failed & 1       & \Failed & 1       & \Failed & 1       & 2       & 63 & \Failed & \Failed & \Failed & \Failed & \Failed & \Failed & \Failed & \Failed \\
4  & 2       & \Failed & 2       & \Failed & 2       & \Failed & 1       & 1       & 34 & 2       & \Failed & 2       & \Failed & 1       & 2       & 1       & 1       & 64 & 1       & \Failed & 1       & 2       & 1       & \Failed & 1       & \Failed \\
5  & 1       & \Failed & 1       & \Failed & 1       & \Failed & 1       & \Failed & 35 & 3       & \Failed & 2       & \Failed & 1       & \Failed & 1       & \Failed & 65 & 1       & \Failed & 1       & \Failed & 1       & \Failed & 1       & \Failed \\
6  & 1       & \Failed & 1       & \Failed & 1       & \Failed & 1       & \Failed & 36 & \Failed & \Failed & \Failed & \Failed & 1       & 2       & 1       & 2       & 66 & 1       & \Failed & 1       & 2       & 1       & 1       & 1       & 1       \\
7  & 1       & \Failed & 1       & \Failed & 1       & \Failed & 1       & \Failed & 37 & 1       & \Failed & 1       & \Failed & 1       & \Failed & 1       & \Failed & 67 & \Failed & \Failed & \Failed & \Failed & \Failed & \Failed & 1       & 1       \\
8  & \Failed & \Failed & \Failed & 2       & \Failed & 2       & \Failed & 2       & 38 & 1       & \Failed & 1       & \Failed & 1       & 1       & 1       & 1       & 68 & \Failed & \Failed & \Failed & \Failed & \Failed & \Failed & \Failed & \Failed \\
9  & 1       & \Failed & 1       & \Failed & 1       & \Failed & 1       & \Failed & 39 & 1       & \Failed & 1       & \Failed & 1       & \Failed & 1       & \Failed & 69 & 1       & \Failed & 1       & \Failed & 1       & \Failed & 1       & 1       \\
10 & 1       & \Failed & 1       & 1       & 1       & 1       & 1       & 1       & 40 & 1       & \Failed & 1       & \Failed & 1       & \Failed & 1       & \Failed & 70 & \Failed & \Failed & \Failed & \Failed & 1       & 1       & 1       & 1       \\
11 & 1       & \Failed & 1       & \Failed & 1       & \Failed & 1       & \Failed & 41 & 1       & \Failed & 1       & \Failed & 1       & \Failed & 1       & 1       & 71 & 1       & \Failed & 1       & \Failed & \Failed & \Failed & \Failed & \Failed \\
12 & \Failed & \Failed & \Failed & 1       & \Failed & 1       & \Failed & 1       & 42 & \Failed & \Failed & \Failed & \Failed & \Failed & \Failed & \Failed & \Failed & 72 & 1       & \Failed & 1       & \Failed & 1       & \Failed & 1       & \Failed \\
13 & 1       & \Failed & 1       & 1       & 1       & 1       & 1       & 1       & 43 & 1       & \Failed & 1       & \Failed & 1       & 1       & 1       & 1       & 73 & 1       & \Failed & 1       & \Failed & 1       & 1       & 1       & \Failed \\
14 & 1       & \Failed & 1       & \Failed & 1       & \Failed & 1       & \Failed & 44 & 1       & \Failed & 1       & \Failed & 1       & \Failed & 1       & \Failed & 74 & 2       & \Failed & 1       & \Failed & 1       & \Failed & 1       & \Failed \\
15 & \Failed & \Failed & \Failed & \Failed & \Failed & \Failed & \Failed & \Failed & 45 & 1       & \Failed & 1       & \Failed & 1       & 1       & 1       & 1       & 75 & 1       & \Failed & 1       & \Failed & 1       & \Failed & 1       & 1       \\
16 & 1       & \Failed & 1       & 2       & 1       & 1       & 1       & 1       & 46 & \Failed & \Failed & \Failed & \Failed & 1       & 1       & 1       & 1       & 76 & 1       & \Failed & 1       & \Failed & 1       & 4       & 1       & 1       \\
17 & \Failed & \Failed & \Failed & \Failed & \Failed & \Failed & \Failed & \Failed & 47 & 1       & \Failed & 1       & \Failed & 1       & 4       & 1       & \Failed & 77 & \Failed & \Failed & \Failed & \Failed & \Failed & 1       & \Failed & 1       \\
18 & \Failed & \Failed & \Failed & \Failed & \Failed & \Failed & \Failed & \Failed & 48 & \Failed & \Failed & \Failed & 8       & \Failed & 8       & \Failed & 8       & 78 & \Failed & \Failed & \Failed & \Failed & \Failed & 1       & \Failed & 1       \\
19 & 1       & \Failed & 1       & \Failed & 1       & \Failed & 1       & \Failed & 49 & \Failed & \Failed & 3       & 2       & 3       & 2       & 3       & 2       & 79 & \Failed & \Failed & \Failed & \Failed & \Failed & \Failed & \Failed & \Failed \\
20 & 1       & \Failed & 1       & \Failed & 1       & \Failed & 1       & \Failed & 50 & 1       & \Failed & 1       & \Failed & 1       & \Failed & 1       & \Failed & 80 & 1       & \Failed & 1       & \Failed & 1       & 1       & 1       & 1       \\
21 & 1       & \Failed & 1       & 1       & 1       & 1       & 1       & 1       & 51 & 1       & \Failed & 1       & \Failed & 1       & \Failed & 1       & \Failed & 81 & 5       & \Failed & 5       & \Failed & 5       & \Failed & 5       & \Failed \\
22 & 1       & \Failed & 1       & 2       & 1       & 2       & 1       & 1       & 52 & 1       & \Failed & 1       & \Failed & 1       & \Failed & 1       & \Failed & 82 & 3       & \Failed & 2       & 1       & 2       & 1       & 2       & 1       \\
23 & 1       & \Failed & 1       & 2       & 1       & \Failed & 1       & \Failed & 53 & 1       & \Failed & 1       & \Failed & 1       & \Failed & 1       & 1       & 83 & \Failed & \Failed & \Failed & \Failed & \Failed & \Failed & \Failed & \Failed \\
24 & 1       & \Failed & 1       & \Failed & 1       & \Failed & 1       & \Failed & 54 & 1       & \Failed & 1       & \Failed & 1       & 1       & 1       & 1       & 84 & 1       & \Failed & 1       & \Failed & 1       & \Failed & 1       & \Failed \\
25 & 1       & \Failed & 1       & \Failed & 1       & \Failed & 1       & \Failed & 55 & \Failed & \Failed & \Failed & \Failed & \Failed & \Failed & \Failed & \Failed & 85 & 1       & \Failed & 1       & \Failed & 1       & \Failed & 1       & \Failed \\
26 & 2       & \Failed & 1       & \Failed & 1       & 1       & 1       & 1       & 56 & 1       & \Failed & 1       & \Failed & 1       & \Failed & 1       & \Failed & 86 & 1       & \Failed & 1       & \Failed & 1       & \Failed & 1       & \Failed \\
27 & 1       & \Failed & 1       & 1       & 1       & 1       & 1       & 1       & 57 & 1       & \Failed & 1       & \Failed & 1       & \Failed & 1       & \Failed & 87 & 1       & \Failed & 1       & \Failed & 1       & 1       & 1       & 1       \\
28 & 1       & \Failed & 1       & 2       & 1       & 2       & 1       & 1       & 58 & 1       & \Failed & 1       & \Failed & 1       & \Failed & 1       & \Failed & 88 & 1       & \Failed & 1       & \Failed & 1       & \Failed & 1       & \Failed \\
29 & 1       & \Failed & 1       & 2       & 1       & 2       & 1       & 2       & 59 & \Failed & \Failed & \Failed & \Failed & \Failed & \Failed & \Failed & \Failed & 89 & \Failed & \Failed & 1       & \Failed & 1       & \Failed & 1       & \Failed \\
30 & 1       & \Failed & 1       & \Failed & 1       & \Failed & 1       & \Failed & 60 & 1       & \Failed & 1       & \Failed & 1       & \Failed & 1       & \Failed &    &         &         &         &         &         &         &         &         \\
    \bottomrule
\end{tabular}
    \end{center}
\end{table*}

\paragraph*{Generalization Ability}

Thanks to the dynamic generation of equation patterns, the rules learned by \EqFix are \emph{insensitive} to where errors locate.
In example \#1 of \cref{tb:motivating}, the erroneous part $``\verb|10|"$ appears at the end of the input equation,
whereas in example \#2, $``\verb|123|"$ appears before $``\verb|+x|"$ in the input equation.
Although their positions are distinct, the rule synthesized by example \#1 is general enough to fix problem \#2,
as the generated equation pattern $P_2$ (in \cref{fig:apply}) replaces $``\verb|123|"$ with $v_1$.
In contrast, the rules learned by \FlashFill are less general---in many cases, they are position-\emph{sensitive} because \FlashFill does not support extracting problem-specific information from error messages,
which is a major difference between the two.
As a result, the rule learned from \#1 by \FlashFill cannot generalize to solve \#2 while \EqFix can.

\paragraph*{Efficiency}

\cref{fig:syn-time} depicts the cumulative synthesis time when the number of training example groups increases under C4.
The average time was \SI{201}{ms} for \EqFix and \SI{476}{ms} for \FlashFill.
\EqFix spent less time on 78 out of 89 runs.

\paragraph*{Relaxer Extension}

We also noticed that in 17 (19.1\%) of the 89 synthesized rules under configuration C4, relaxers exhibited in the synthesized rule.
This reveals that the relaxer extension is necessary and improves our tool's practicality.

\paragraph*{Failure Cases}

\EqFix failed on 17 test cases under C4.
Manually inspecting these cases, we classified the cause of failure into three categories:
\begin{itemize}
    \item Inconsistent examples (7 cases):
    The provided examples are \emph{inconsistent} with each other, so our synthesizer failed to give any consistent rule.
    \ifExt
    To overcome this failure, the user has to either modify these examples by hand or suggest alternative consistent ones.
    \fi
    \item Insufficient error message (6 cases):
    The provided error messages are \emph{insufficient}, and \EqFix could not generate a useful equation pattern.
    \ifExt
    For example, the error message ``\verb|Missing $ inserted|'' of \#12 does not contain any problem-specific information.
    Usually, the entire input equation must be transformed, and since our string transformation DSL is inadequately expressive, the learned rules are overfitting the training examples.
    \fi
    \item Restricted DSL expressiveness (4 cases):
    The test case is \emph{deviated far} from the training examples of the same group, and due to the restricted expressiveness of our DSL, the learned rule could not generalize to that test case.
\end{itemize}

\paragraph*{Testing the Rule Library}

Additionally, we conducted another evaluation on the same dataset, but only for \EqFix, that fitted a more realistic setting where equation repair problems are solved by trying the initial rules saved in a rule library.
We obtained the initial rule library by learning from the entire training set examples (C4) under the training mode.
Then, under the applying mode, we tested all 89 test cases.
Interestingly, compared with the results shown in \cref{tb:exp1}, one more test case (\#71) was solved (by the top-ranked rule synthesized from the example group \#69).

\section{Related Work}\label{sec:related}

\paragraph*{Program Repair}
Automated program repair aims to automatically correct programs 
so that they satisfy the desired specification \cite{Challenges-Repair}.
Heuristic-based repair tools such as \tool{GenProg}
\cite{GenProg-GP,GenProg,GenProg-Study}
employs an extended form of genetic programming with heuristics\zap{and evolves program variants until the correct one is found}.
However, it is shown that these techniques produce patches that overfit the test suite \cite{Overfitting}.
Ranking techniques have been studied to address the problem.
\tool{ACS} \cite{ACS} produces precise patches\zap{that have a relatively high probability to be correct}
with a refined ranking technique for condition synthesis.
\tool{PAR} \cite{PAR} mines bug fix patterns from the history
and gives frequently occurring fixes high priority\zap{in the random search process}.
\tool{Prophet} \cite{PatchGen} outperforms the previous works by
learning a probabilistic model for ranking the candidate patches.

Semantics-based repair techniques generate repairs via symbolic execution \cite{SemFix,Angelix} and program synthesis \cite{SPR}.
Such techniques, however, are also suspected of overfitting the test suite.
Recently, a new repair synthesis engine called \tool{S3} is proposed \cite{S3}.
It leverages \emph{Programming by Examples} (PBE) methodology to synthesize high-quality bug repairs, elaborating several ranking features.

Technically, \EqFix belongs to the semantics-based family.
To avoid overfitting, we also rely on the ranking technique for promoting rules with a high generality.
Furthermore, syntactic errors are common and must be tackled in equation repair, while it is usually neglected in program repair, as people concern more about bugs \cite{Repair-Survey}.
\tool{HelpMeOut} \cite{error-messages} aids developers to debug compilation and run-time error messages by suggesting past solutions.
Unlike \EqFix, it only provides related examples and cannot create repairs for new problems.

\NoFAQ \cite{NoFAQ} is a system for fixing buggy Unix commands using PBE.
Although the problem domain is similar to ours, due to its lazy synthesis algorithm,
it only can synthesize a practical fixing rule when at least two examples are provided.
In contrast, in many cases, one example is sufficient for \EqFix.
The DSL of \NoFAQ can only accept tokenized strings (separated by spaces) as input.
However, there is no direct way to tokenize an erroneous equation in our problem domain.
We thus introduce the equation patterns and propose a mechanism for synthesizing them.
These equation patterns help to pattern match against an equation and extract the variant parts, which need to be transformed later.

\paragraph*{Text Transformation}
\FlashFill \cite{flash-fill} pioneered in text transformation via program synthesis and was later extended for semantic transformation \cite{Semantic-String}.
A similar technique is put into a live programming environment by \tool{StriSynth} \cite{StriSynth}.
\tool{FlashExtract} \cite{flash-extract} automates data extraction by highlighting texts on web pages.
String transformation is performed at a high level in these techniques, but it is unsuitable for repairing equations.
We realize that error messages guide the repairing process.
By pattern matching the error messages, we only perform the transformation on a few variables instead of the entire equation, which takes less time.

\paragraph*{VSA-based Program Synthesis}
\emph{Version space algebra} (VSA) has been widely adopted in PBE applications
\cite{VSA-PBD,flash-normalize,flash-meta,ringer,transformation,prog-trans}.
In those applications, VSA is critical as the set of candidate programs is possibly very large.
In \EqFix, the synthesized rules are represented compactly with VSA.

VSA-based program synthesis has also been applied in program transformation.
\tool{Refazer} \cite{prog-trans} is a framework that automatically learns program transformations
at an abstract syntax tree level.
Feser et al. \citet{data-structure} propose a method for example-guided synthesis of
recursive data structure transformations in functional programming languages.
Nguyen et al. \citet{graph-based} present a graph-based technique that guides developers in adapting API usages.
\section{Conclusion \& Future Directions}\label{sec:conclusion}

We present \EqFix, a system for fixing both compilation and typesetting errors in \LaTeX{} equations.
We design a DSL for expressing fixing rules and propose a synthesis algorithm to learn rules from user-provided examples.
In the future, our tool can be improved by leveraging data from various sources like \LaTeX{} online forums via crowdsourced learning.
When a large rule library is established, it would be interesting to develop an \EqFix plugin in modern editors for practical use.
Furthermore, because adding more data does not require any change in \EqFix but simply needs more input-output examples to construct synthesis rules, our approach can potentially be applied to other string and mathematical equation systems.

\bibliographystyle{splncs04}
\bibliography{synthesis,repair}

\begin{thebibliography}{10}
\providecommand{\url}[1]{\texttt{#1}}
\providecommand{\urlprefix}{URL }
\providecommand{\doi}[1]{https://doi.org/#1}

\bibitem{ringer}
Barman, S., Chasins, S., Bodik, R., Gulwani, S.: Ringer: web automation by
  demonstration. In: Proceedings of the 2016 ACM SIGPLAN International
  Conference on Object-Oriented Programming, Systems, Languages, and
  Applications. pp. 748--764. ACM (2016)

\bibitem{NoFAQ}
D'Antoni, L., Singh, R., Vaughn, M.: {NoFAQ}: Synthesizing command repairs from
  examples. In: Proceedings of the 2017 11th Joint Meeting on Foundations of
  Software Engineering. pp. 582--592. ESEC/FSE 2017, ACM, New York, NY, USA
  (2017). \doi{10.1145/3106237.3106241}

\bibitem{data-structure}
Feser, J.K., Chaudhuri, S., Dillig, I.: Synthesizing data structure
  transformations from input-output examples. In: Proceedings of the 36th ACM
  SIGPLAN Conference on Programming Language Design and Implementation. pp.
  229--239. PLDI '15, ACM, New York, NY, USA (2015).
  \doi{10.1145/2737924.2737977}

\bibitem{GenProg}
Goues, C.L., Nguyen, T., Forrest, S., Weimer, W.: {GenProg}: A generic method
  for automatic software repair. IEEE Transactions on Software Engineering
  \textbf{38}(1),  54--72 (Jan 2012). \doi{10.1109/TSE.2011.104}

\bibitem{Challenges-Repair}
Goues, C., Forrest, S., Weimer, W.: Current challenges in automatic software
  repair. Software Quality Journal  \textbf{21}(3),  421--443 (Sep 2013).
  \doi{10.1007/s11219-013-9208-0}

\bibitem{StriSynth}
Gulwani, S., Mayer, M., Niksic, F., Piskac, R.: {StriSynth}: Synthesis for live
  programming. In: 2015 IEEE/ACM 37th IEEE International Conference on Software
  Engineering. vol.~2, pp. 701--704 (May 2015). \doi{10.1109/ICSE.2015.227}

\bibitem{flash-fill}
Gulwani, S.: Automating string processing in spreadsheets using input-output
  examples. In: Proceedings of the 38th Annual ACM SIGPLAN-SIGACT Symposium on
  Principles of Programming Languages. pp. 317--330. POPL '11, ACM (2011).
  \doi{10.1145/1926385.1926423}

\bibitem{pbe}
Gulwani, S., Esparza, J., Grumberg, O., Sickert, S.: Programming by examples
  (and its applications in data wrangling). Verification and Synthesis of
  Correct and Secure Systems  (2016)

\bibitem{error-messages}
Hartmann, B., MacDougall, D., Brandt, J., Klemmer, S.R.: What would other
  programmers do: Suggesting solutions to error messages. In: Proceedings of
  the SIGCHI Conference on Human Factors in Computing Systems. pp. 1019--1028.
  CHI '10, ACM, New York, NY, USA (2010). \doi{10.1145/1753326.1753478}

\bibitem{flash-normalize}
Kini, D., Gulwani, S.: {FlashNormalize}: Programming by examples for text
  normalization. In: Proceedings of the 24th International Conference on
  Artificial Intelligence. pp. 776--783. AAAI Press (2015)

\bibitem{VSA-PBD}
Lau, T.A., Domingos, P., Weld, D.S.: Version space algebra and its application
  to programming by demonstration. In: Proceedings of the Seventeenth
  International Conference on Machine Learning. pp. 527--534. ICML '00, Morgan
  Kaufmann Publishers Inc., San Francisco, CA, USA (2000)

\bibitem{flash-extract}
Le, V., Gulwani, S.: {FlashExtract}: A framework for data extraction by
  examples. In: Proceedings of the 35th ACM SIGPLAN Conference on Programming
  Language Design and Implementation. pp. 542--553. PLDI '14, ACM (2014).
  \doi{10.1145/2594291.2594333}

\bibitem{PAR}
Le, X.B.D., Lo, D., Goues, C.L.: History driven program repair. In: 2016 IEEE
  23rd International Conference on Software Analysis, Evolution, and
  Reengineering (SANER). vol.~1, pp. 213--224 (March 2016).
  \doi{10.1109/SANER.2016.76}

\bibitem{S3}
Le, X.B.D., Chu, D.H., Lo, D., Le~Goues, C., Visser, W.: S3: Syntax- and
  semantic-guided repair synthesis via programming by examples. In: Proceedings
  of the 2017 11th Joint Meeting on Foundations of Software Engineering. pp.
  593--604. ESEC/FSE 2017, ACM, New York, NY, USA (2017).
  \doi{10.1145/3106237.3106309}

\bibitem{GenProg-Study}
Le~Goues, C., Dewey-Vogt, M., Forrest, S., Weimer, W.: A systematic study of
  automated program repair: Fixing 55 out of 105 bugs for \$8 each. In:
  Proceedings of the 34th International Conference on Software Engineering. pp.
  3--13. ICSE '12, IEEE Press, Piscataway, NJ, USA (2012)

\bibitem{SPR}
Long, F., Rinard, M.: Staged program repair with condition synthesis. In:
  Proceedings of the 2015 10th Joint Meeting on Foundations of Software
  Engineering. pp. 166--178. ESEC/FSE 2015, ACM, New York, NY, USA (2015).
  \doi{10.1145/2786805.2786811}

\bibitem{PatchGen}
Long, F., Rinard, M.: Automatic patch generation by learning correct code. In:
  Proceedings of the 43rd Annual ACM SIGPLAN-SIGACT Symposium on Principles of
  Programming Languages. pp. 298--312. POPL '16, ACM, New York, NY, USA (2016).
  \doi{10.1145/2837614.2837617}

\bibitem{Angelix}
Mechtaev, S., Yi, J., Roychoudhury, A.: Angelix: Scalable multiline program
  patch synthesis via symbolic analysis. In: Proceedings of the 38th
  International Conference on Software Engineering. pp. 691--701. ICSE '16,
  ACM, New York, NY, USA (2016). \doi{10.1145/2884781.2884807}

\bibitem{Repair-Survey}
Monperrus, M.: Automatic software repair: A bibliography. ACM Comput. Surv.
  \textbf{51}(1),  17:1--17:24 (Jan 2018). \doi{10.1145/3105906}

\bibitem{graph-based}
Nguyen, H.A., Nguyen, T.T., Wilson, Jr., G., Nguyen, A.T., Kim, M., Nguyen,
  T.N.: A graph-based approach to {API} usage adaptation. In: Proceedings of
  the ACM International Conference on Object Oriented Programming Systems
  Languages and Applications. pp. 302--321. OOPSLA '10, ACM, New York, NY, USA
  (2010). \doi{10.1145/1869459.1869486}

\bibitem{SemFix}
Nguyen, H.D.T., Qi, D., Roychoudhury, A., Chandra, S.: {SemFix}: Program repair
  via semantic analysis. In: Proceedings of the 2013 International Conference
  on Software Engineering. pp. 772--781. ICSE '13, IEEE Press, Piscataway, NJ,
  USA (2013)

\bibitem{flash-meta}
Polozov, O., Gulwani, S.: {FlashMeta}: A framework for inductive program
  synthesis. ACM SIGPLAN Notices  \textbf{50}(10),  107--126 (2015)

\bibitem{prog-trans}
Rolim, R., Soares, G., D'Antoni, L., Polozov, O., Gulwani, S., Gheyi, R.,
  Suzuki, R., Hartmann, B.: Learning syntactic program transformations from
  examples. In: Proceedings of the 39th International Conference on Software
  Engineering. pp. 404--415. ICSE '17, IEEE Press, Piscataway, NJ, USA (2017).
  \doi{10.1109/ICSE.2017.44}

\bibitem{Semantic-String}
Singh, R., Gulwani, S.: Learning semantic string transformations from examples.
  Proc. VLDB Endow.  \textbf{5}(8),  740--751 (Apr 2012).
  \doi{10.14778/2212351.2212356}

\bibitem{Overfitting}
Smith, E.K., Barr, E.T., Le~Goues, C., Brun, Y.: Is the cure worse than the
  disease? overfitting in automated program repair. In: Proceedings of the 2015
  10th Joint Meeting on Foundations of Software Engineering. pp. 532--543.
  ESEC/FSE 2015, ACM, New York, NY, USA (2015). \doi{10.1145/2786805.2786825}

\bibitem{GenProg-GP}
Weimer, W., Nguyen, T., Goues, C.L., Forrest, S.: Automatically finding patches
  using genetic programming. In: 2009 IEEE 31st International Conference on
  Software Engineering. pp. 364--374 (May 2009).
  \doi{10.1109/ICSE.2009.5070536}

\bibitem{ACS}
Xiong, Y., Wang, J., Yan, R., Zhang, J., Han, S., Huang, G., Zhang, L.: Precise
  condition synthesis for program repair. In: Proceedings of the 39th
  International Conference on Software Engineering. pp. 416--426. ICSE '17,
  IEEE Press, Piscataway, NJ, USA (2017). \doi{10.1109/ICSE.2017.45}

\bibitem{transformation}
Yaghmazadeh, N., Klinger, C., Dillig, I., Chaudhuri, S.: Synthesizing
  transformations on hierarchically structured data. In: Proceedings of the
  37th ACM SIGPLAN Conference on Programming Language Design and
  Implementation. pp. 508--521. PLDI '16, ACM (2016).
  \doi{10.1145/2908080.2908088}

\end{thebibliography}


\end{document}